\documentclass[useAMS,usenatbib,usegraphicx]{mn2e}
\title[]{A search for extragalactic H$_2$O maser emission towards IRAS galaxies II -- Discovery of an H$_2$O maser in the Seyfert 1 galaxy NGC\,4051}
\author[]{Yoshiaki Hagiwara,$^{1}$\thanks{E-mail: hagiwara@astron.nl (YH)} Philip J Diamond,$^{2}$ Makoto Miyoshi,$^{3}$  Emmanouel Rovilos,$^{2}$   \and
Willem Baan$^{1}$ \\ 
$^{1}$ASTRON, Westerbork Observatory, P.O. Box 2, Dwingeloo, 7990 AA, The Netherlands\\
$^{2}$Jodrell Bank Observatory, University of Manchester, Macclesfield, Cheshire, SK11 9DL, UK\\
$^{3}$National Astronomical Observatory, Osawa 2-21-1, Mitaka, Tokyo, Japan}
\begin{document}
\newcommand{\km}{km s$^{-1}\;$}
\newcommand{\kms}{km s$^{-1}\;$}
\newcommand{\kmss}{km s$^{-1}$}
\newcommand{\cmt}{cm$^{-3}$}
\newcommand{\vlsr}{{\it V}$_{\rm LSR}\;$}
\newcommand{\vs}{V$_{\rm sys}$}
\newcommand{\mb}{mJy  beam$^{-1}$}
\newcommand{\msun}{\mbox{M$_{\sun}$}}
\newcommand{\lsun}{\mbox{L$_{\sun}$}}
\newcommand{\lf}{\mbox{L$_{\rm FIR}$}}
\newcommand{\flb}{\mbox{S$_{\rm 21 cm}$}}
\newcommand{\ho}{H$_{2}$O$\;$}
\newcommand{\hho}{H$_{2}$O}
\date{Accepted 2003 July 9. Received 2003 June 18; in original form 2003 March 4}

\pagerange{\pageref{}--\pageref{}} \pubyear{2003}

\maketitle

\label{}

\begin{abstract}
Water vapor maser emission in the 6$_{16}$--5$_{23}$ transition
towards the narrow-line Seyfert 1 (NLS1) galaxy NGC\,4051 has been
discovered during an ongoing single-dish extragalactic water maser
survey. The Doppler-shifted maser components appear to
bracket maser components lying near the systemic velocity of the
galaxy symmetrically. The tentative result of a Very Large Array~(VLA) snapshot observation is that the
masers are confined within 0.1 arcsec (5 pc at a distance D = 9.7 Mpc)
of the radio continuum peak seen at 8.4~GHz. The low luminosity of 
the maser ($\sim$ 2 \lsun) is not typical for masers that
coincide with the radio continuum nucleus and appear associated with Active Galactic Nucleus~(AGN) activity. A low-luminosity maser in a Type 1
Seyfert nucleus could be explained by a low maser gain resulting from the lower inclination of an obscuring disk around an active nucleus.
\end{abstract}


\section[]{Introduction}
The existence of super massive black holes in active galactic nuclei
(AGN) was clearly proved by the detection of a thin Keplerian rotating
disk in the LINER galaxy NGC\,4258 that was imaged using strong \ho
maser emission \citep{miyo95}. VLBI observations have shown that these
masers are distributed in the thin disk with a radius of 0.13 -- 0.26
pc tracing a region down to $\sim$ 40,000 Schwarzschild radii, which
is the lowest radial distance ever imaged in the 'central engine' of
an AGN \citep{herr96}. Motivated by this discovery, a number of
single-dish surveys for new water masers in AGN have been carried out
(e.g., \citealt{braa96}) and have yielded a catalogue of about 30
sources in early 2003 (\citealt{linc03}). Studies of several \ho
masers at milliarcsecond resolution using VLBI have shown that the
masers do not always lie around an active nucleus \citep{mora99}, but that some of
them are associated with other AGN activities such as jets or nuclear
outflows. \ho masers are found to be important tools for investigating the nuclear kinematics in various types of active galaxies, in particular
narrow-line AGN. A number of broad-line AGN have been surveyed for
masers, however, no water maser emission has yet been detected in
broad-line Seyfert 1 galaxies. In 2001, we began a survey program
searching for \ho masers in far-infrared (FIR) luminous galaxies
(\citealt{hagi02}, hereafter Paper I). In Paper I we reported the
first confirmed detection of a maser in the Seyfert galaxy
NGC\,6240. The aim of this survey was to find new \ho masers in
`radio-excess' FIR luminous galaxies, which were selected on the basis
of a large ratio of the correlated flux density at 21-cm (using the
VLA at $2-25$~arcsec resolution) to the FIR flux density. Details on
the sample selection were described in Paper I. The new detection in
NGC\,6240 motivated us to continue the survey of the 'radio-excess'
FIR galaxies. \\
In this letter, we briefly summarize the results of the ongoing maser
survey following Paper I, which has culminated in the discovery of a
water maser in the Type 1 Seyfert galaxy NGC\,4051. In addition, we
present preliminary results from VLA snapshot observations of this
maser.
\section[]{The survey}
Survey observations of \ho emission in the 6$_{16}$-5$_{23}$
transition at 22.23508 GHz have been carried out with the MPIfR 100-m
radio telescope at Effelsberg in March and September 2002. Throughout
the survey, a K-band HEMT receiver with two orthogonal linear
polarizations was used with a system temperature of 70--120 K
depending on atmospheric conditions. The spectrometer was an 8-channel
digital autocorrelator (AK 90), each channel having a bandwidth of 40
MHz ($\sim$ 540 \kmss) and 512 spectral points (velocity resolution
$\sim$ 1.1 \kmss). Four channels were employed for each
polarization. The centre spectral points of each correlator were
offset from the systemic velocity of the galaxies by typically $\pm$
(20 -- 30) MHz, thereby providing a total bandwidth of 100 -- 130 MHz
per polarization.  More details on the observations can be found in
Paper I.
\section[]{Survey results}
A total of 24 galaxies from the FIR galaxy sample have been observed,
including previously observed objects in this survey. Ten newly
observed galaxies are listed in Table 1. During the survey, we
discovered a new water maser toward a narrow-line Seyfert 1 (NLS1)
galaxy NGC\,4051 \citep{oste85} on 5 March 2002. The detection was
confirmed six months later on 14 September. The parameters of the
galaxy are summarized in Table 2, where the systemic velocity of the
galaxy (\vlsr = 730 $\pm$ 5 \kmss) is adopted. The sample of 24
selected galaxies consists of Type 1 and 2 Seyferts, compact HII
nuclei and LINERs. The two new detections in NGC\,6240 and NGC\,4051
suggest an $\sim$~8\% detection rate, which is greater than that of
$\sim$~3\% in \citet{braa96} and that of $\sim$~4\% in \citet{linc03},
all of which are based on modest number statistics. While the galaxies
surveyed in \citet{braa96} and \citet{linc03} were selected from a
distance-limited AGN sample with systemic velocities $\la$ 8000 \kmss,
we constructed a radio-excess FIR galaxy sample based on the radio and
FIR properties of the AGN.  The total number of radio-excess FIR
galaxies is about 30, of which at present eight galaxies contain water
masers. These results suggest that the presence of the \ho masers in
active galaxies correlates well with the combined presence of a considerable
FIR flux density (at 60~$\umu$m and 100~$\umu$m) and a radio compact core
originating in the dust heated by an active nucleus or a compact HII
region \citep{baan96}. \\
The detection of an \ho water maser in Type 1 Seyfert galaxies is
still rather unique since most of the extragalactic \ho masers have
been found in Type 2 Seyferts or LINERs. Since the Seyfert galaxy
NGC\,5506, which contains an \ho maser, has recently been
re-classified as a NLS1, NGC\,4051 is the second detection in this
type of galaxy \citep{naga02}. Figure 1 shows the spectrum of the
maser in NGC\,4051 observed at two epochs 6 months apart. The emission
is dominated by narrow components with linewidths of $\sim$ 1 -- 2
\kmss. Several components in the range \vlsr = $\sim$ 670 -- 690 \kms
and at \vlsr = 712 \kms varied by a factor of two over six months. The
spectrum in Figure 1 shows that several Doppler-shifted maser
components are seen on either side of the systemic velocity. Their
centroid velocities are listed in Table 3.

A first interpretation is that the maser components are distributed
quasi-symmetrically relative to the 730 \kms component at the systemic
velocity.  The discovery spectrum also displays a high-velocity
component at \vlsr = 936 \kms ($\sim$ 3 $\sigma$ level), but this
component does not have any blue-shifted counterparts. The maser
components are observed from 659 \kms to the weaker component at 773
\kmss, which accounts for a velocity extreme of $\pm$ 70 \kms relative
to the systemic velocity. While these extrema are not as large as the
maximum rotation velocity of $\approx $ 1100 \kms for NGC\,4258 (e.g.,
\citealt{miyo95}), such symmetrically or quasi-symmetrically
distributed emission has not been observed except in the cases of
NGC\,4258, NGC\,1068 and NGC\,2960 \citep{linc96,henk02}. We searched
for additional components within $\pm$ 1000 \kms of the systemic
velocity, but none were detected at the 3 $\sigma$ sensitivity levels of
15 -- 45 mJy, depending on the velocity range (Table 1).  No
acceleration of the maser components was detected over the six months
interval, yielding an upper limit for the drift rate of $\sim$ 2.2
\kms year$^{-1}$.  The total maser luminosity, assuming isotropic
radiation of the emission, is $\approx $ 2 \lsun. The galaxy has been
searched for water masers at least twice, in 1983 \citep{clau84} and
1995 \citep{braa96}, with a 3 $\sigma$ detection level of 130 -- 180
mJy. These observations were not sensitive enough to be able to detect
even the strongest maser component of $\sim $ 90 mJy shown in Figure
1. Therefore, it is uncertain whether the detected components represent a
flaring state for the galaxy. However, the galaxy could be a promising
object for study of a 'classical' accretion disk around an active
nucleus if the maser components are indeed associated with AGN
activity.
\begin{table*}
 \centering
 \begin{minipage}{140mm}
  \caption{List of galaxies observed with the Effelsberg 100m telescope during 2002. The coordinates of each source are from the NED database.}
  \begin{tabular}{@{}lllllll@{}}
  \hline 
~~Source &~~$\alpha$  &~~$\delta$ &~~\vs\footnote{Primarily from NED database}  &$\Delta$V\footnote{Observed velocity range} &Rms noise&Epoch\footnote{Observing epoch; A(4--6 March), B(14--15 September), both in 2002} \\
&(B1950)&(B1950)&(\km)&(\km)&(mJy)& \\ 
 \hline
 
NGC 3227  &10$^h$ 20$^m$ 43$^s$&+20$\degr$ 09$\arcmin$ 06$\arcsec$&1157&260 -- 1960&10&A \\
NGC 3367  &10 43 56&+14 00 51&3037&2100 -- 3700&10&A\\
NGC 4051  &12 00 36&+44 48 35&730&200 -- 1550&5&A\\
          &12 00 36&+44 48 35&730&-300 -- 1800&15&B\\
IRAS 12112+0305  &12 11 12&+03 05 20&21980&21070 -- 22370&12&A \\
NGC 4418  &12 24 20&-00 36 09 &2179&1550 -- 2860&20&A\\
NGC 4594  &12 37 23&-11 20 55&1091&490 -- 1770&15&A\\
IC 883 &13 18 17&+34 24 05&7000&6260 -- 7570&12&A\\
NGC 5256 &13 36 14&+48 31 45&8211&7400 -- 9120 &10&B\\
PKS1345+12 &13 45 06 &+12 32 20&36575&35900 -- 37180&10&A\\
ZW 49.057 &15 10 45&+07 24 37&3897&3280 -- 4980&13&A\\
\hline
\end{tabular}
\end{minipage}

\end{table*}
\section[]{VLA observation of NGC\,4051}
On 26 April 2002, a snapshot observation was made with the NRAO
\footnote{The National Radio Astronomy Observatory is a facility of
the National Science Foundation operated under cooperative agreement
by Associated Universities, Inc.} Very Large Array (VLA) in A
configuration to determine the positions and spatial distribution of
the \ho masers in NGC\,4051. We employed two 6.25 MHz IFs,
corresponding to a velocity coverage of 84 \kmss, that were centered
on \vlsr = 680 \kms and 730 \kms to cover all known velocity
components in the range \vlsr = 645 -- 770 \kmss. Each IF was divided
into 64 spectral channels, yielding a velocity resolution of 1.3
\kmss. The observation was made in phase-referencing mode using a
nearby phase-calibrator J\,1153+495. 3C\,286 was used for amplitude
and bandpass calibration. The duration of the observation was 2 hours,
resulting in an rms noise of 0.4 \mb per spectral channel after
smoothing the channel spacing to 2.6 \kmss. The uniformly weighted
synthesized beam (FWHM) was 0.12 arcsec $\times$ 0.076 arcsec (P.A. =
73$^{\circ}$). No continuum emission was detected in this observation
to a 3 $\sigma$ level of $\sim$ 0.3 \mb. Several velocity components
were tentatively detected at or less than the 3 $\sigma$ detection
level. The positions of most of these components were found to lie in
a $\sim$ 0.04~arcsec region centered on Right Ascension (J2000) =
12$^{\rm h}$03$^{\rm m}$09$^{\rm s}$.61, Declination (J2000)=
+44$^{\circ}$31$\arcmin$52$\arcsec$.6. The components peaking at \vlsr
= 712 \kms and 740 \kms were detected at the $\sim$ 3 $\sigma$ level,
both components were unresolved by the VLA beam. Their
beam-deconvolved sizes are typically $\sim$ 60 mas $\times$ 40 mas,
corresponding to $\sim$ 3 pc $\times$ 2 pc in linear scale. Other
velocity components were marginally detected with 2 -- 2.5 $\sigma$
levels, but we will not discuss the detection of these
components. Further sensitive observations are necessary to measure
the precise positions of all the components mentioned above.
%
%
\begin{table}
  \caption{Properties of NGC 4051. Velocity adopted from de Vaucouleurs et al. (1991) is converted to heliocentric systemic velocity using the radio definition. Optical inclination and the distance come from Adams (1977). The radio flux density at 21-cm was measured by the VLA in A configuration (Ulvestad \& Wilson 1984). Far-infrared flux densities and luminosity are from the NED data base and Condon et al. (1990). The X-ray luminosity is from Collinge et al. (2001). We assume that $H_0$ = 75 km s$^{-1}$ Mpc$^{\;-1}$.}
  \begin{tabular}{@{}ll@{}}
  \hline 
Systemic Velocity (21-cm HI)  & 723 $\pm$ 5   km s$^{-1}$ (Heliocentric) \\
~~~~~~~~~~~~~~~~~~~~~~~~~           & 730 $\pm$ 5   km s$^{-1}$ (LSR) \\
(optical)&  686 $\pm$ 13  km s$^{-1}$  \\
Distance                &  9.7 Mpc \\
Inclination (optical)  & 40$\degr$ \\
Optical class    & Seyfert 1 \\
F$_{\nu}$ (20-cm) & 19 $\pm$ 1.5  mJy \\
F$_{\nu}$ (100-$\umu$m)& 23.9 $\pm$ 0.1 Jy\\
F$_{\nu}$ (60-$\umu$m) & 10.4 $\pm$ 0.1 Jy\\
F$_{\nu}$ (25-$\umu$m) & 2.2 $\pm$ 0.1 Jy\\
L$_{\rm FIR}$ (40 -- 120 $\umu$m) & 10$^{9.57}$ \lsun  \\
L$_{\rm Xray}$ (2 -- 10 keV) &  (2.3--5.7) 10$^{41}$ erg s$^{-1}$ \\                                                  
\hline
\end{tabular}
\end{table}
\section[]{The \ho maser in NGC\,4051}
The position of the radio continuum emission at 8.4~GHz (R.A.(J2000) =
12$^{h}$03$^{m}$09$^{s}$.605, Dec.(J2000) =
+44$^{\circ}$31$\arcmin$52$\arcsec$.73) was measured with 0.3~arcsec
resolution of the VLA in A-configuration \citep{kuku95}. The
astrometric accuracy of this position was estimated to be between 0.01
and 0.05 arcsec in \citet{kuku95}. Assuming that the frequency dependence of the radio
continuum position is negligible, the position of the masers is
coincident with that of the continuum emission within astrometric
uncertainties of $\approx$ 0.1 arcsec, or 5 pc at the adopted distance
of 9.7 Mpc. The tentatively detected maser components therefore lie in
the vicinity of the radio nucleus of NGC\,4051. There is also a
suggestion of weaker components with a small spatial offset from the
radio continuum peak, but this needs to be confirmed.

A question remains as to the true origin of the maser in NGC\,4051. Is
it associated with a star-forming region or AGN activity? The presence
of the marginally detected high-velocity component, which is
red-shifted by $\approx$ 200 \kmss, hints that the origin of the maser
is AGN activity. The most important point of our discovery is that the
maser lies in a Type 1 Seyfert nucleus. NLS1 galaxies exhibit
broad-line optical emission, characteristic of Type 1 Seyfert nuclei,
together with narrow-line emission; except that these broad-line
widths are much narrower ($<$ 2000 \kmss) than in typical Type 1
nuclei (e.g., \citealt{oste85}). Observers are likely to look at only
a part of the Broad Line Region (BLR) in the nucleus of such galaxies.
It has been argued that low-luminosity (L$_{\rm {H_2O}}$ $\la$ 10 \lsun) masers are apt to be found outside of active nuclei \citep{clau86} but the origin of such a maser in a Type 1 Seyfert nucleus may require a different explanation.
%
%
%
\begin{table}
  \caption{List of velocities (peak velocities in the flux density) of each maser component close to the systemic velocity obtained from Fig~1.  Uncertainties of velocities are 0.5 --  1.1 \kmss. The errors of the flux density are typically 10 percent.}
  \begin{tabular}{llll}
 \hline
\multicolumn{2}{c}{MAR 4-5}&\multicolumn{2}{c}{SEP 14-15}  \\
\hline
\vlsr&Flux density&\vlsr&Flux density\\
(\kmss)&(mJy)&(\kmss)&(mJy) \\ 
  \hline 
   659 & 15 &  &    \\
   664 & 15&664& 30  \\
    674 & 43&674& 32\\
    679 & 54&679& 44  \\
 683 &70 &    & \\
 696 & 9&    &\\
 701 & 17&    &  \\
 712 & 38&712&85 \\
 730 & 19&730& 24\\
 738 & 43&   & \\
 741 & 86&741& 85\\
 773 & 20&  &  \\
 936 & 14&  &  \\
\hline
\end{tabular}
\end{table}
\section[]{Interpretation of the low-luminosity}
The low-luminosity ($\approx$ 2 \lsun) of the 'nuclear' water maser
and the modest Doppler shifts of any maser components in the nucleus
of NGC\,4051 can be explained by a lower inclination angle of the disk
or disk-like configuration in a heavily obscured circumnuclear medium
along our line of sight (LOS) \citep{chri97}. This configuration could
significantly decrease the maser gain along the LOS, which results in
a lower maser luminosity, as compared with the extragalactic water
masers with L$_{\rm {H_2O}}$ $>$ 100 \lsun\, associated with a highly
inclined disk plane. Peterson et al. (2000) argued that the Balmer
lines in NGC\,4051 might arise in a low-inclination disk-like
configuration with an inclination similar to 
the 40$^{\circ}$ of the galaxy as a whole, 
which we would preferentially observe on the near
side. They also found that the narrow-line objects could be located at
the same position as the AGN BLR. This implies that the \ho emitting
medium would coincide with the structure that obscures the BLR.  The
putative lower disk-inclination model is consistent with the observed
smaller LOS velocities ({\it v}$_{\rm los}$ = {\it v}$_{\rm rad}$ sin$i$, $i$ =
disk inclination angle) of the high-velocity components in a disk. The
lower galactic disk inclination (30$^{\circ}$-- 40$^{\circ}$) in NGC\,4051 would reduce the
apparent LOS velocities by about 40 -- 50 per cent as compared to
those in a more  edge-on disk ($i$ $>$ 70$^{\circ}$). 

The galaxy shows strong and rapid X-ray variations \citep{lawr87} but
there is no direct evidence yet for strong maser variability. There is
no prominent starburst region in the galaxy and the total galaxy
mass-to-FIR luminosity ratio is 10$^3$ times less than that of typical
Galactic starforming regions like W\,51 \citep{smit83}, which makes a
relation with active starformation implausible. We cannot still rule out
that the maser in NGC\,4051 is associated with starforming activity on
the basis of high-resolution observations with pc-scale
resolution. However, we favor the idea that AGN activity gives rise to
the \ho maser in the galaxy since NGC\,4051 contains an AGN and the
physical environment could prevent a high-luminosity maser.
The \ho maser in the NLS1 galaxy NGC\,5506 shows several narrow (1-2
\kmss) maser lines near the systemic velocity with variable maser flux
densities of 0.1-0.3 Jy \citep{braa96}. The luminosity of such spiky
components is on the order of 1 \lsun, which is comparable with those
in NGC\,4051. The similarity of the NGC\,4051 maser to those in
NGC\,5506 is also evident in the broad variable components (line
widths of 20-80 \kmss) near the systemic velocities, which are found
in most of the \ho masers associated with AGN activity. The relative
weakness of the masers in both galaxies probably results from the
environment surrounding these active nuclei, because narrow optical
lines and hard X-rays would correlate with intense \ho masers
\citep{oste85}. Compton reflection components in the form of neutral
iron lines in hard X-ray bands have been detected in both NGC\,5506 and
NGC\,4051 \citep{matt01}. This could support the presence of an
optically thick neutral layer shrouding the Type 1 Seyfert nucleus and
the BLR. The detection of an \ho maser in an NLS1 would also tell us
about unified models of AGN because the \ho masers may trace the medium
that obscures the BLR in Type 1 Seyfert nuclei.  A complete sample of
NLS1 objects should be searched for new \ho masers using sensitive
telescopes.

\section*{Acknowledgments}

This research has made use of the NASA/IPAC Extragalactic Database 
(NED) which is operated by the Jet Propulsion Laboratory, California
Institute of Technology, under contract with the National Aeronautics
and Space Administration. YH appreciates the efforts of Barry Clark for
his immediate scheduling of VLA observing time.

\newpage
\begin{figure}
\includegraphics[angle=0,height=11cm,width=8cm]{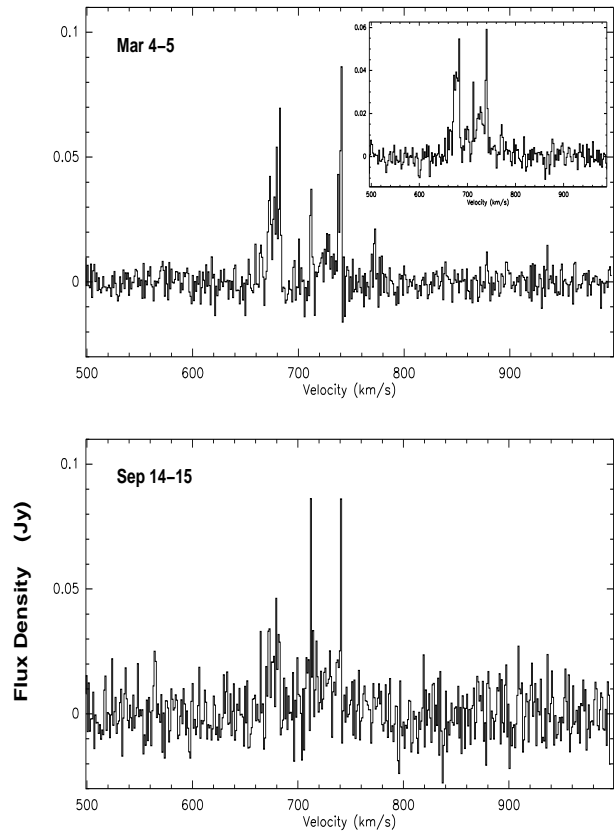}       
\caption{Spectra of the \ho maser in  NGC\,4051, obtained with the MPIfR 100 m telescope at two epochs March and September 2002. The velocity resolution is 1.1 \kmss. The adopted systemic velocity of NGC\,4051 is \vlsr = 730 $\pm$ 5 \kmss. The velocities in the spectra are scaled in the radio LSR convention. The spectra were produced by averaging the data taken at two observing days for each epoch.  A velocity-smoothed spectrum is displayed as an inset with the resolution of 2.1 \kmss.}
\end{figure}

\bsp

\label{lastpage}

\end{document}